\documentclass[conference]{IEEEtran}

\IEEEoverridecommandlockouts
\usepackage{graphicx}

\usepackage{amsmath,mathtools}
\usepackage{amssymb}
\usepackage{bbm}
\usepackage{xcolor}
\usepackage{tabularx}
\usepackage{algorithm}
\usepackage{algpseudocode}
\algrenewcommand\algorithmicrequire{\textbf{Input:}}
\algrenewcommand\algorithmicensure{\textbf{Output:}}
\usepackage{tikz}
\usetikzlibrary{arrows.meta, positioning, shapes.multipart}
\usepackage{geometry}
\usetikzlibrary{decorations.pathreplacing}
\def\BibTeX{{\rm B\kern-.05em{\sc i\kern-.025em b}\kern-.08em
    T\kern-.1667em\lower.7ex\hbox{E}\kern-.125emX}}
\usepackage{url}
\usepackage{cite}
\newcommand{\llrrparen}[1]{% \llrrparen{..}
\left(\mkern-3mu\left(#1\right)\mkern-3mu\right)}
\begin{document}
% figure and ADFM legacy for candidate

%%%%%%
\title{AFDM Channel Parameter Estimation: Exploiting its Shift Properties and Input-Output Relationship \vspace{-0.35cm}}
\author{\IEEEauthorblockN{Jialiang~Zhu\textsuperscript{*},\;
Zeping~Sui\textsuperscript{$\dagger$},\;
Zilong~Liu\textsuperscript{$\dagger$},\;
Arman~Farhang\textsuperscript{*}
} 
\IEEEauthorblockA{\textsuperscript{*} Department of Electronic \& Electrical Engineering, Trinity College Dublin, Ireland 
\\ \textsuperscript{$\dagger$} School of Computer Science and Electronic Engineering, University of Essex, UK
\\Email: zhuj3@tcd.ie, zepingsui@outlook.com, zilong.liu@essex.ac.uk, arman.farhang@tcd.ie
}
\vspace{-3em}
\thanks{
The work of Jialiang Zhu and Arman Farhang was supported by Research Ireland grants 18/CRT/6222 and 24/US/4013 and the by the MULTIPLY-6G project with the Grant Agreement No. 101293106 funded under the European Union’s Horizon Europe research and innovation programme. The work of Zilong Liu and Zeping Sui was supported in part by the U.K. Engineering and Physical Sciences Research Council (EPSRC) under Grants EP/Y037243/1, EP/X040569/1, EP/X035352/1, and EP/Y000986/1, and by the Royal Society under Grants IEC$\backslash$NSFC$\backslash$233292 and IES$\backslash$R1$\backslash$241212. For the purpose of Open Access, the author has applied a CC BY public copyright licence to any Author Accepted Manuscript version arising from this submission..
}
 }

%%%%%%%

\maketitle

\begin{abstract}
Affine frequency division multiplexing (AFDM) has attracted significant attention owing to its excellent backwards compatibility, full diversity achievability, as well as strong resilience to high Doppler and various hardware impairments. Existing AFDM channel parameter estimation methods, however, either require large pilot overhead or rely on computationally intensive grid search methods. This paper proposes a grid-search-independent, low-overhead, impulse-pilot-based channel parameter estimation method for AFDM systems by exploiting its inherent shift properties. Specifically, we show that the effect of a Doppler shift on the impulse pilot is equivalent to a delay shift in the time domain with a phase rotation. Additionally, both delay and Doppler shifts lead to cyclic shifts in the affine domain. By utilizing these properties in both the time and affine domains, two linear equations are formed to estimate the integer delay and Doppler of each path. The fractional part of the Doppler is estimated from the Doppler-induced leakage around the impulse pilot by exploiting the AFDM input-output relationship. For multipath channels, a successive interference cancellation (SIC) framework is developed, in which the strongest propagation path is first estimated, reconstructed, and canceled before the remaining paths are successively estimated. Simulation results demonstrate that the proposed method significantly reduces computational complexity and allows lower pilot overhead than the benchmarks. The proposed method also achieves improved channel parameter estimation and full channel estimation accuracy, with lower bit error rate (BER).

\end{abstract}


\vspace{-0.2cm}
\section{Introduction}
Future sixth-generation (6G) wireless communication systems are expected to support reliable high-mobility communications and integrated sensing and communications (ISAC) over highly dynamic wireless channels in diverse application scenarios, such as unmanned aerial vehicles (UAVs), high-speed trains, and low-Earth-orbit (LEO) satellite communications \cite{3gpp}. In these applications, 
conventional orthogonal frequency-division multiplexing (OFDM) struggles due to high inter-carrier interference incurred by large Doppler and rapidly varying channels. 

Against this context, affine frequency division multiplexing (AFDM) has emerged recently as a promising waveform candidate for high-mobility channels \cite{afdm,afdm_sui,sui2025multi}. Based on the discrete affine Fourier transform (DAFT), AFDM establishes an affine delay-Doppler representation of the wireless channel, where each propagation path is transformed into a shift in the affine domain \cite{sui2026mimo}. By appropriately selecting the post-chirp parameter $c_1$, the effective channel entries corresponding to different delay-Doppler pairs may occupy distinct positions in the affine domain \cite{afdm}.
Consequently, AFDM enables unambiguous identification of the associated channel parameters.

As a fundamental requirement for every coherent communication system, channel estimation directly determines the performance of subsequent data detection. However, existing AFDM channel parameter estimation methods generally suffer from either high pilot overhead or high computational complexity. The original AFDM channel estimation scheme in~\cite{afdm_journal} employed an impulse pilot embedded with guard symbols. Although this method provides reliable channel parameter estimation, the large guard region introduces a significant pilot overhead. To this end, the authors in~\cite{afdm_gi} proposed a guard-free impulse-pilot scheme. A superimposed pilot design that eliminates the guard region was also conceived in~\cite{afdm_super}. Whilst both approaches can reduce pilot overhead, the pilot power is generally required to be more than 20 dB higher than that of the data symbols to ensure reliable channel estimation. As a result, less amount of the transmit power is allocated to the data part, thus degrading the overall data detection performance. To reduce the estimation complexity, the basis expansion model (BEM)-assisted channel estimation for AFDM was proposed in~\cite{afdm_bem}. However, their method requires two impulse pilots, leading to doubled guard overhead. An off-grid sparse Bayesian learning (SBL) method in~\cite{afdm_sbl} performed an Doppler refinement following the conventional on-grid delay-Doppler search. Nonetheless, the expectation-maximization (EM) required by the SBL refinement substantially increase the computational complexity.

In view of the these challenges in AFDM channel estimation, the contributions of this paper are summarized as follows:
\begin{itemize}
    \item We introduce a channel parameter estimation technique using an impulse pilot for AFDM systems. The proposed method exploits the inherent delay and Doppler shift properties of AFDM in time and affine domain, enabling efficient estimation of the integer delay and Doppler.

    \item We develop a fractional Doppler estimation technique by exploiting the Doppler-induced leakage around the impulse pilot based on the AFDM input-output relationship. 
    Since the proposed technique is grid-search independent, it achieves significantly lower computational complexity.

    \item For multipath channels, we introduce an iterative successive interference cancellation (SIC) framework, in which the strongest propagation path is first estimated, reconstructed, and canceled. The same procedure is then applied successively  to estimate the remaining paths.

\end{itemize}

Simulation results demonstrate that the proposed channel parameter estimation technique achieves better channel estimation accuracy, and improved BER performance than peak selection and the orthogonal matching pursuit (OMP)-based benchmark, and allows less pilot overhead.

\textit{Notation:}
Scalars, vectors, and matrices are denoted by regular lowercase, bold lowercase, and bold uppercase letters, respectively.
The superscripts $(\cdot)^{\rm{H}}$ and $(\cdot)^{\rm{T}}$ indicate Hermitian and transpose operations, respectively. 
$\mathbf{I}_{N}$ is the identity matrix of size $N$ and $\mathbf{0}_{M\times N}$ is $M\times N$ matrices consisting of zeros.
$\delta[\cdot]$ is the Dirac delta function. $\llrrparen{\cdot}_N$ is the modulo $N$ operator. 

\section{Affine Frequency Division Multiplexing}
%\vspace{-0.1cm}
In AFDM, complex-valued data symbols are placed in the DAFT domain.
The DAFT-domain symbols are then mapped to the time domain via the inverse DAFT (IDAFT). 
Considering $N$ chirp subcarriers and the vector $\mathbf{x}$
containing the independent and identically
distributed (i.i.d.) quadrature amplitude modulated (QAM) data symbols with zero mean and unit power, the time domain signal is represented as 
\begin{equation}
    \mathbf{s}
    =
    \mathbf{A}^{\mathrm{H}}\mathbf{x}
    ,
    \label{eq:afdm_modulation_matrix}
\end{equation}
where $\mathbf{A}=\boldsymbol{\Lambda}_{c_2}\mathbf{F}_N\boldsymbol{\Lambda}_{c_1}$ is the DAFT matrix and $c_1,c_2$ are post-chirp and pre-chirp parameters, respectively. Moreover, $\boldsymbol{\Lambda}_{c}=\operatorname{diag}\{e^{-j2\pi c n^{2}}, n=0, 1, ..., N-1\}$ is the diagonal chirp matrix with the chirp parameter $c$ and $\mathbf{F}_N$ is the normalized $N$-point discrete Fourier transform (DFT) matrix.
To mitigate inter-symbol interference caused by the multipath channel, a chirp-periodic prefix (CPP) is prepended to each AFDM symbol. The resulting signal can be expressed as
\begin{equation}
    \mathbf{s}_{\rm cpp}=\mathbf{A}_{\rm cpp}\mathbf{s},
    \label{eqn:s_cpp}
\end{equation}
where $\mathbf{A}_{\rm cpp}=[(\mathbf{\Omega}{\mathbf G}_{\rm cpp})^{\rm T}, {\mathbf{I}_{N}^{\rm T}}]^{\rm T}$is the CPP addition matrix. Here, $\mathbf{\Omega}={\rm{diag}} \{e^{-j2\pi c_1(N^2+ 2N\ell)},\ell=-L_\text{cpp},\ldots, -1\}$ and $L_\text{cpp}$ is the CPP length, which is selected to exceed the delay spread. The ${L}_{\rm cpp} \times N$ matrix $\mathbf{G}_{\rm cpp}$ is constructed from the last ${L}_{\rm cpp}$ rows of the identity matrix $\mathbf{I}_{N}$.

After the AFDM signal passes through the time-varying channel, comprising $P$ propagation paths, the received signal is given by
\begin{equation}
    \mathbf{r}_{\rm cpp}
    =
    \mathbf{H}\mathbf{s_{\rm cpp}}
    +
    \mathbf{w},
    \label{eq:time_domain_channel}
\end{equation}
where $\mathbf{H}=\sum_{p=0}^{P-1}h_p \widetilde{\boldsymbol{\Delta}}^{\nu_p}\boldsymbol{L}_{l_p}$ is the time domain channel matrix with $h_p$, $\nu_p$ and $l_p$ denoting the complex-valued path gain, normalized Doppler shift and normalized delay, respectively, of path $p$. $\widetilde{\boldsymbol{\Delta}}^{\nu_p}= \operatorname{diag}\{e^{\frac{j2\pi\nu_p}{N}n}, n=-L_{\rm cpp}, \ldots, N-1\}$  represents Doppler-induced phase rotation by $\nu_p$. The square matrix $\boldsymbol{L}_{l_p}$, with elements $L_{l_p}[m,n]=\delta[{m-(n+l_p)}]$ for $m,n=0\ldots,N+L_{\rm cpp}-1$, represents a linear time shift with $l_p$ samples. Finally, $\mathbf{w}    \sim
    \mathcal{CN}
    \left(
        \mathbf{0},
        \sigma_w^2\mathbf{I}_N
    \right)$ is the complex additive white Gaussian noise (AWGN) with the variance $\sigma_w^2$.

At the receiver, the CPP is first removed, i.~e.,
\begin{equation}
    \mathbf{r}
    =
    \mathbf{R}_{\rm cpp}\mathbf{r}_{\rm cpp}=
    \overline{\mathbf{H}}\mathbf{s}
    +
    \overline{\mathbf{w}},
    \label{eq:time_domain_channel}
\end{equation}
where $\mathbf{R}_{\rm cpp} = [\mathbf{0}_{N \times {L_{\rm cpp}}}, \mathbf{I}_{N}]$ is the CPP removal matrix.

The resulting channel matrix after CPP removal is given by $\overline{\mathbf{H}}=\mathbf{R}_{\rm cpp}\mathbf{H}\mathbf{A}_{\rm cpp}=\sum_{p=0}^{P-1}
    h_p\boldsymbol{\Gamma}_{\mathrm{cpp}_p}
{\boldsymbol{\Delta}}^{\nu_p}
    \boldsymbol{\Pi}^{l_p}$ where 
\begin{align}
\boldsymbol{\Gamma}_{\mathrm{CPP}_p}
&\!=\!
\operatorname{diag}
\left(
\left\{
\begin{array}{ll}
e^{-j2\pi c_1\left(N^2-2N(l_p-n)\right)}, & n<l_p,\\[2mm]
1, & n\ge l_p,
\end{array}
\right.
\right),
\end{align}
The Doppler shift matrix for the $i^{\rm th}$ path is $\mathbf{\Delta}^{\nu_i}= \operatorname{diag}\{e^{\frac{j2\pi\nu_i}{N}n}, n=0, \ldots, N-1\}$ and $\boldsymbol{\Pi}$ denotes the unit cyclic shift matrix with elements $\Pi[m,n]=\delta[m-(\!(n+1)\!)_N]$ for $m,n=0,\ldots,N-1$. The noise vector after CPP removal is given by $\overline{\mathbf{w}}=\mathbf{R}_{\rm cpp}\mathbf{w}$. 
The received signal is then transformed into the affine domain using the DAFT as
\begin{equation}
    \mathbf{y}
    =
    \mathbf{A}\mathbf{r}
    =
   \mathbf{A}
    \overline{\mathbf{H}}
    \mathbf{A}^{\mathrm{H}}
    \mathbf{x}
    +
    \mathbf{A}\overline{\mathbf{w}}
    = \mathbf{H}_{\mathrm{eff}}\mathbf{x}
    +
    \widetilde{\mathbf{w}},
    \label{eq:afdm_demodulation}
\end{equation}
where $\mathbf{H}_{\mathrm{eff}}
    =\mathbf{A}\overline{\mathbf{H}}\mathbf{A}^{\mathrm{H}}$ and $    \widetilde{\mathbf{w}}
    =
    \mathbf{A}\overline{\mathbf{w}}$
represent the effective channel and the noise term, respectively.
Since $\mathbf{A}$ is a unitary matrix, the statistical properties and noise power remain the same in the affine domain, i.~e., 
    $\widetilde{\mathbf{w}}
    \sim
    \mathcal{CN}
    \left(
        \mathbf{0},
        \sigma_w^2\mathbf{I}_N
    \right).$
Finally, the transmit data symbols can be estimated using linear minimum mean square error (LMMSE) equalization, i.e.,
\begin{equation}
    \widehat{\mathbf{x}}
    =
    \left(
        \mathbf{H}_{\mathrm{eff}}^{\mathrm{H}}
        \mathbf{H}_{\mathrm{eff}}
        +
        {\sigma_w^2}\mathbf{I}_N
    \right)^{-1}
    \mathbf{H}_{\mathrm{eff}}^{\mathrm{H}}
    \mathbf{y}.
    \label{eq:afdm_lmmse}
\end{equation}

\section{Time and Frequency Shift Effects on AFDM}
\label{sec:3}
In this section, we investigate the effects of delay and Doppler shift on AFDM. For an affine domain impulse pilot, a Doppler-induced frequency shift appears as a shift in the time domain with phase rotation which is an interesting observation. Additionally, 
a timing offset results in a shift in the affine domain with a phase rotation. 
These properties lay the foundation for our proposed channel parameter estimation technique in Section~\ref{sec:4}.
\subsection{Effect of Doppler shift in time domain}
The discrete time-domain AFDM signal can be written as
\begin{equation}
 s[n]=\frac{1}{\sqrt N}\sum_{m=0}^{N-1} x[m]\,
e^{j2\pi\left(c_1 n^2+\frac{nm}{N}+c_2 m^2\right)}, n=0,\dots,N-1.   
\end{equation}
Assume the transmitted signal $\mathbf{x}_{\rm p}$ contains only impulse pilot at subcarrier index \(m_{\rm p}\), i.e. $x_{\rm p}[m]=\sqrt{\rho}\,\delta[m-m_{\rm p}]$,
where $\rho$ is the power of the impulse pilot. Then, the  time-domain impulse pilot becomes
\begin{equation}
    s_{\rm p}[n]=A\,e^{j\pi\frac{ \mu n^2}{N}}e^{j2\pi\frac{ m_{\rm p} n}{N}},
    \label{eq:s_p}
\end{equation}
where $\mu = 2Nc_1$ and $A={\sqrt {\rho N}}e^{j2\pi c_2 m_{\rm p}^2}$.
Under the condition of a normalized Doppler shift $\varepsilon$ is applied to $s_{\rm p}[n]$, therefore,
\begin{equation}
s_1[n]=s_{\rm p}[n]e^{j2\pi\frac{ \varepsilon n}{N}}=A\,e^{j\pi\frac{ \mu n^2}{N}}\,e^{j2\pi\frac{ (m_{\rm p} + \varepsilon)n}{N}}.
\label{s2}
\end{equation}

Assume ${s}_2[n]$ is a $d_1$-sample shift version of $s_{\rm p}[n]$, then %$s_2[n]$ can be formulated as 
\begin{align}
s_2[n]\!&=\!s_{\rm p}[\llrrparen{n-d_1}_N]=A\,e^{j\pi\frac{\mu (n-d_1)^2}{N}}\,e^{j2\pi\frac{ m_{\rm p}(n-d_1)}{N}}\nonumber\\
\!&=\!
A\,e^{-j2\pi\frac{d_1}{N}(m_{\rm p}-\frac{\mu d_1}{2})}\,
e^{j\pi\frac{ \mu n^2}{N}}\,
e^{j2\pi\frac{ (m_{\rm p}-\mu d_1)n}{{N}}}.
\label{s1}
\end{align}

It can be observed that except a constant phase term $e^{-j2\pi\frac{d_1}{N}(m_{\rm p}-\frac{\mu d_1}{2})}$ in (\ref{s1}), (\ref{s2}) and (\ref{s1}) have the same form. Therefore, $s_1[n]$ is equivalent to $s_{\rm p}[n]$ with a constant phase and a circular shift by $d_1$ samples when
$\llrrparen{m_{\rm p}+\varepsilon}_N\equiv \llrrparen{m_{\rm p}-\mu d_1}_N$. We can further obtain 
\begin{equation}
\llrrparen{\varepsilon}_N \equiv \llrrparen{- \mu d_1}_N.
\label{eq:mod}
\end{equation}
Hence, the time-domain Doppler-affected impulse pilot signal $s_1[n]$ can be seen as a circularly shifted version of the impulse pilot signal $s_{\rm p}[n]$ with a constant phase, 
%which can be expressed as
\begin{equation}
s_1[n]
=
e^{j2\pi\frac{d_1}{N}(m_{\rm p}-\frac{\mu d_1}{2})}\,
s_{\rm p}[\llrrparen{n-d_1}_N].
\end{equation}

\subsection{Effect of timing offset in affine domain}
We first recall the expression of the time-domain impulse pilot with a $d_1$-sample shift given in (\ref{s1}).
Let us transform $s_2[n]$ into the affine domain, we have
\begin{align}
S_2[k]
\!\!&=\!\!
\frac{1}{\sqrt N}\sum_{n=0}^{N-1}
A' e^{j2\pi c_1 n^2}
e^{j2\pi \frac{m_{\rm p}-\mu d_1}{N}n}
e^{-j2\pi\left(c_1 n^2+\frac{nk}{N}+c_2 k^2\right)}\nonumber\\
&=\frac{A'}{\sqrt N}
e^{-j2\pi c_2 k^2}
\sum_{n=0}^{N-1}
e^{j2\pi \frac{n(m_{\rm p}-\mu d_1-k)}{N}},
\label{eq:Sk}
\end{align}
where $A'=
A\,e^{-j2\pi\frac{d_1}{N}(m_{\rm p}-\frac{\mu d_1}{2})}.$
Based on the DFT orthogonality, we can derive
$
\sum_{n=0}^{N-1}
e^{j2\pi \frac{n(m_{\rm p}-\mu d_1-k)}{N}}
=
N\,\delta[k-\llrrparen{m_{\rm p}-\mu d_1}_N].
$
Therefore, (\ref{eq:Sk}) can be reformulated as 
\begin{align}
S_2[k]&=
A'\sqrt N\,
e^{-j2\pi c_2 k^2}\,
\delta[k-\llrrparen{m_{\rm p}-\mu d_1}_N].
\end{align}
The only non-zero value of $S_2[k]$ is located at $k=\llrrparen{m_{\rm p}-\mu d_1}_N$
, which shows that the pilot still holds the impulse property with the location of $\llrrparen{m_{\rm p}-\mu d_2}_N$ and its complex amplitude is imposed with an extra phase factor.

Therefore, a delay and a Doppler shift produce shifts in both the time and affine domains. 
A path delay results in a circular displacement of the impulse pilot in the affine domain and a delay-dependent phase rotation. 
Similarly, a Doppler shift leads to a circular shift in the time domain and a Doppler-dependent phase rotation. 
These relationships are closely related to the shift properties of linear canonical transform (LCT), where time translations and frequency modulations correspond to coordinate displacements and phase rotations. This highlights the underlying LCT structure of the chirp-based AFDM waveform\cite{shi2014generalized}.

\section{Proposed Channel Parameter Estimation}
\label{sec:4}
In this section, we propose a channel parameter estimation technique using an impulse pilot. 
The proposed technique first identifies the strongest propagation path by estimating its integer delay and Doppler shifts. The fractional Doppler shift is then estimated, followed by estimation of the corresponding path gain by solving a least-squares (LS) problem. The contribution of the identified path is subsequently reconstructed and subtracted from the received signal. The procedure is repeated on the residual signal until all path parameters are found\footnote{Assuming a sufficiently high sampling rate, the path delays are modeled as integer-valued while Doppler shifts consist of integer and fractional parts. We further assume that the number of propagation paths is known a priori.}.

\begin{figure}[t]
\centering
\scalebox{0.65}{
\begin{tikzpicture}[font=\sffamily]

% Parameters
\def\nboxes{20} % total boxes per row
\def\boxsize{0.6}
\def\dataStart{6} % data starts at this box
\def\dataEnd{15}  % data ends at this box
\def\leakEnd{9} % data starts at this box
\def\leakStart{13}  % data ends at this box
\def\leakEndd{16} % data starts at this box
\def\leakStartt{12}  % data ends at this box
% Colors
\definecolor{pilot}{RGB}{255,220,120}
\definecolor{guard}{RGB}{255,245,210}
\definecolor{data}{RGB}{255,170,100}

% ----------- BS row -----------------
\foreach \i in {1,...,\nboxes}{
    \ifnum\i=18
        \node[draw,fill=data,minimum size=\boxsize cm] (bs\i) at (\i*\boxsize,0) {...};
    \else\ifnum\i=11
        \node[draw,fill=guard,minimum size=\boxsize cm] (bs\i) at (\i*\boxsize,0) {...};
    \else\ifnum\i=4
        \node[draw,fill=data,minimum size=\boxsize cm] (bs\i) at (\i*\boxsize,0) {...};
    % \else\ifnum\i=8
    %     \node[draw,fill=data,minimum size=\boxsize cm] (bs\i) at (\i*\boxsize,0) {...};

    \else\ifnum\i=1
        \node[draw,fill=data,minimum size=\boxsize cm] (bs\i) at (\i*\boxsize,0) {$\sqrt{\rho}$};    
    \else\ifnum\i<4
        \node[draw,fill=data,minimum size=\boxsize cm] (bs\i) at (\i*\boxsize,0) {0};
    \else\ifnum\i<7
        \node[draw,fill=data,minimum size=\boxsize cm] (bs\i) at (\i*\boxsize,0) {0};
    \else\ifnum\i<11
        \node[draw,fill=guard,minimum size=\boxsize cm] (bs\i) at (\i*\boxsize,0) {$d$};
    \else\ifnum\i<16
        \node[draw,fill=guard,minimum size=\boxsize cm] (bs\i) at (\i*\boxsize,0) {$d$};
    \else\ifnum\i<18
        \node[draw,fill=data,minimum size=\boxsize cm] (bs\i) at (\i*\boxsize,0) {0};    
    \else
        \node[draw,fill=data,minimum size=\boxsize cm] (bs\i) at (\i*\boxsize,0) {0};
    \fi\fi\fi\fi\fi\fi\fi\fi\fi
}
% \foreach \i in {1,...,\nboxes}{
%     \ifnum\i=19
%         \node[draw,fill=guard,minimum size=\boxsize cm] (bs\i) at (\i*\boxsize,0) {...};
%     \else\ifnum\i=14
%         \node[draw,fill=data,minimum size=\boxsize cm] (bs\i) at (\i*\boxsize,0) {...};
%     \else\ifnum\i=3
%         \node[draw,fill=guard,minimum size=\boxsize cm] (bs\i) at (\i*\boxsize,0) {...};
%     \else\ifnum\i=8
%         \node[draw,fill=data,minimum size=\boxsize cm] (bs\i) at (\i*\boxsize,0) {...};

%     \else\ifnum\i=11
%         \node[draw,fill=data,minimum size=\boxsize cm] (bs\i) at (\i*\boxsize,0) {$P$};    
%     \else\ifnum\i<3
%         \node[draw,fill=guard,minimum size=\boxsize cm] (bs\i) at (\i*\boxsize,0) {$d$};
%     \else\ifnum\i<6
%         \node[draw,fill=guard,minimum size=\boxsize cm] (bs\i) at (\i*\boxsize,0) {$d$};
%     \else\ifnum\i<11
%         \node[draw,fill=data,minimum size=\boxsize cm] (bs\i) at (\i*\boxsize,0) {0};
%     \else\ifnum\i<17
%         \node[draw,fill=data,minimum size=\boxsize cm] (bs\i) at (\i*\boxsize,0) {0};
%     \else\ifnum\i<19
%         \node[draw,fill=guard,minimum size=\boxsize cm] (bs\i) at (\i*\boxsize,0) {$d$};    
%     \else
%         \node[draw,fill=guard,minimum size=\boxsize cm] (bs\i) at (\i*\boxsize,0) {$d$};
%     \fi\fi\fi\fi\fi\fi\fi\fi\fi\fi
% }
\draw[decorate,decoration={brace,amplitude=5pt}] (bs1.north west) -- (bs6.north east);
\draw[decorate,decoration={brace,mirror, amplitude=5pt}] (bs2.south west) -- (bs6.south east) node[midway,yshift=-12pt]{$Q/2$};
\draw[decorate,decoration={brace, amplitude=5pt}] (bs16.north west) -- (bs20.north east) ;
% \draw[decorate,decoration={brace,amplitude=5pt}] (bs1.north west) -- (bs5.north east) ;
% \draw[decorate,decoration={brace,amplitude=5pt}] (bs17.north west) -- (bs21.north east) ;
%\draw[decorate,decoration={brace,mirror, amplitude=5pt}] (bs14.south west) -- (bs16.south east) node[midway,yshift=-12pt]{$G_{data}$};
\draw[decorate,decoration={brace,mirror, amplitude=5pt}] (bs16.south west) -- (bs20.south east) node[midway,yshift=-12pt]{$Q/2$};
\draw[decorate,decoration={brace,mirror,amplitude=5pt}] (bs7.south west) -- (bs15.south east) node[midway,yshift=-12pt]{$\mathcal{D}$};
\draw[->] ([yshift=5pt]bs4.north west) -- ([yshift=34pt]bs11.west);
\draw[->] ([yshift=5pt]bs18.north) -- ([yshift=34pt]bs11.east);
% \draw[->] ([yshift=5pt]bs1.north) -- ([yshift=40pt]bs11.west);
\node at ([yshift=30pt]bs11.north) {$\mathcal{P}$};

\end{tikzpicture}
}
\vspace{-1em}
\caption{AFDM frame design\cite{afdm_journal}.}
\vspace{-0.5em}
\label{fig:guard}
\end{figure}
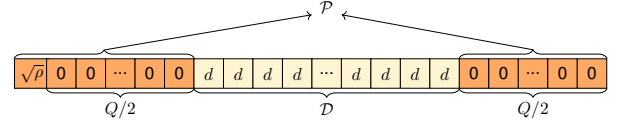
The AFDM frame design from \cite{afdm_journal} is used as shown in Fig.~\ref{fig:guard}. The AFDM subcarrier grid is partitioned into pilot and data regions, defined by the mutually exclusive index sets $\mathcal{P}$ and $\mathcal{D}$ for pilot and data, respectively. 
To ensure unit transmit signal power, the pilot power is assigned as ${\rho}=Q+1$. 

\subsection{Integer Delay and Doppler Estimation}\label{subsec:IntDD}
Our proposed channel estimation finds the path parameters in an iterative manner where at each iteration, we find the parameters for the dominant path and the rest of the paths are absorbed to noise. Consider estimating the $i^{\rm th}$ path which is characterized by delay $l_i$ and Doppler shift $\nu_i=\alpha_i+a_i$, where $\alpha_i$ is the integer part of the normalized Doppler shift and $a_i$ is the fractional part of the Doppler shift with $-0.5<a_i\leq0.5$. When an impulse pilot passes through the channel, Doppler shift leads to a delay shift that is combined with the path delay in time and then the same effect is also observed in the affine domain. Therefore, computing the cross-correlation between the received signal $\mathbf{r}$ and the transmitted pilot signal $\mathbf{s}_{\rm p}$, the dominant correlation peak occurs at the sample index $n_{1,i}$, corresponding to the strongest propagation path. The correlation function is given by $c[n]=\left| \left\langle \mathbf{r}, \mathbf{s}_{\rm p}^{(n)} \right\rangle \right|$, where $\mathbf{s}_{\rm p}^{(n)}$ denotes the circularly shifted version of $\mathbf{s}_{\rm p}$ by $n$ samples. 
According to the shift properties derived in Section \ref{sec:3}, the time-domain location of the peak can be expressed as
\begin{equation}
    n_{1,i}
    =
    \llrrparen{
        l_i+d_i
    }_N, 
    \label{eq:n1}
\end{equation}
where $d_i$ is the time-domain shift induced by Doppler shift $\alpha_i$. We can further obtain $\llrrparen{d_i}_N=\llrrparen{n_{1,i}-l_i}_N$ and if we substitute $d_i$ into (\ref{eq:mod}), we have 
\begin{equation}
\llrrparen{\alpha_i}_N = \llrrparen{- \mu \llrrparen{n_{1,i}-l_i}_N}_N.
\label{eq:affine_shift}
\end{equation}
Applying the DAFT to the received signal $\mathbf{r}$, the peak corresponding to the strongest path can be observed at location $n_{2,i}$ in the affine domain, given by
\begin{equation}
    n_{2,i}
    =
    \llrrparen{
        m_{\rm p}
        -\mu l_i
        +\alpha_i
    }_N.
    \label{eq:fa_mapping}
\end{equation}
The integer delay and Doppler parameters are jointly estimated using \eqref{eq:affine_shift} and \eqref{eq:fa_mapping}. Specifically, the
integer delay candidate \(l_i\) is searched over the range
\(0 \leq l_i \leq l_{\max}\), where \(l_{\max}\) is maximum delay shift. For each delay candidate, the corresponding Doppler index \(\alpha_i\) is computed using \eqref{eq:affine_shift}. The candidate pair
\((\widehat{l}_i,\widehat{\alpha}_i)\) is then accepted if it satisfies
\eqref{eq:fa_mapping}.
Unlike conventional estimation methods, \cite{afdm_super} and \cite{10963873}, which rely solely on the unambiguous affine-domain location of each delay-Doppler pair, the proposed method jointly exploits the shift properties in both the time and affine domains. As a result, the proposed technique remains applicable even
when the delay-Doppler-induced affine-domain shift exceeds the AFDM block length \(N\) and wraps around modulo \(N\).

\subsection{Fractional Doppler Estimation}\label{subsec:Frac_Doppler}
Considering AFDM input-output relationship from \cite{afdm_journal}, the elements of the vector $\mathbf{y}$ in (\ref{eq:afdm_demodulation}) are given by 
\begin{align}
y[k]
&=
\sum_{p=0}^{P-1}
\frac{1}{N}
h_p e^{j2\pi c_1 l_p^2}
\!\!\sum_{q=(k+\operatorname{loc}_p-k_{\nu})N}
     ^{(k+\operatorname{loc}_p+k_{\nu})N}
\!\!\!\!\!\!e^{j\frac{2\pi}{N}
\left(-q l_p+N c_2\left(q^2-k^2\right)\right)}\nonumber
\\
&\quad\times
\frac{
e^{j2\pi\left(k-q+a_p+\operatorname{loc}_p\right)}-1
}{
e^{j\frac{2\pi}{N}
\left(k-q+a_p+\operatorname{loc}_p\right)}-1
}
x[q]
+\widetilde{w}[k], 
\end{align}
for $k=0,\ldots , N-1$, where $\operatorname{loc}_p
\triangleq\llrrparen{\alpha_p + 2N c_1 l_p}_N$ denotes the approximate location of the received pilot from the path $p$, considering the transmit pilot at $m_{\rm p}=0$, i.~e., $x[q]=\sqrt{\rho}\delta[q]$. 
Treating contributions from all paths except the strongest path $i$ as interference,
the input-output relationship reduces to
\begin{equation}
y[k]
=
\frac{\sqrt{\rho}}{N}h_i
e^{j2\pi\left(c_1 l_i^2-c_2 k^2\right)}
\frac{
e^{j2\pi\left(k+a_i+\operatorname{loc}_i\right)}-1
}{
e^{j\frac{2\pi}{N}\left(k+a_i+\operatorname{loc}_i\right)}-1
}
+\eta[k],
\label{eq:IO_simple}
\end{equation}
where $\eta[k]=\sum_{\substack{p=0 \\ p\neq i}}^{P-1}\frac{\sqrt{\rho}}{N}h_p
e^{j2\pi\left(c_1 l_p^2-c_2 k^2\right)}
\frac{
e^{j2\pi\left(k+a_p+\operatorname{loc}_p\right)}-1
}{
e^{j\frac{2\pi}{N}\left(k+a_p+\operatorname{loc}_p\right)}-1
}
+\widetilde{w}[k]$ contains interference from other paths and channel noise. 
%Hence, the only unknown in (\ref{eq:IO_simple}) is $a_i$. 
For simplicity of the following derivation, the noise and interference term, $\eta[k]$, is neglected. Nevertheless, both noise and interference are taken into account in the simulation results. 
Using
$\left|e^{j\theta}-1\right|=
2\left|\sin\left(\frac{\theta}{2}\right)\right|$ and $\left|
\sin\left(\pi\left(k+a_i+\operatorname{loc}_i\right)\right)
\right|=\left|\sin(\pi a_i)\right|$ as $(k+\operatorname{loc}_i)$ is an an integer, the absolute value of $y[k]$ can be approximated as
\begin{equation}
|y[k]|
=
\frac{|h_i|\sqrt{\rho}}{N}
\frac{
\left|\sin(\pi a_i)\right|
}{
\left|
\sin\left(
\frac{\pi}{N}
\left(k+a_i+\operatorname{loc}_i\right)
\right)
\right|
}.
\label{eq:yp}
\end{equation}
Since the integer delay and Doppler are already estimated in the previous section, $\widehat{\operatorname{loc}}_i=\llrrparen{\widehat{\alpha}_i + 2N c_1 \widehat{l}_i}_N$ is obtained. Considering $k_{\rm c}\!=\!\llrrparen{-\widehat{\operatorname{loc}}_i}_N$,
we have
$\llrrparen{k_{\rm c}\!+\!\widehat{\operatorname{loc}}_i}_N\!\!=\!0$. Since 
$\left|
\sin\left(\frac{\pi a_i}{N}\right)
\right|
=
\sin\left(\frac{\pi|a_i|}{N}\right)$ for $a_i\in(-0.5,0.5]$, 
\begin{equation}
\left|
y\!\left[k_{\rm c}\right]
\right|
=
\frac{|h_i|\sqrt{\rho}}{N}
\frac{|\sin(\pi a_i)|}
{\sin\left(\frac{\pi|a_i|}{N}\right)}.
\end{equation}
For the left neighbor of $k_{\rm c}$,
$
k_{\rm l}=\llrrparen{-\widehat{\operatorname{loc}}_i-1}_N,
$
we obtain
\begin{equation}
\left|
y\!\left[k_{\rm l}\right]
\right|
=
\frac{|h_i|\sqrt{\rho}}{N}
\frac{|\sin(\pi a_i)|}
{\sin\left(\frac{\pi}{N}(1-a_i)\right)}.
\end{equation}
Similarly, for the right neighbor of $k_{\rm c}$,
$
k_{\rm r}=\left(-\widehat{\operatorname{loc}}_i+1\right)_N,
$
\begin{equation}
\left|
y\!\left[k_{\rm r}\right]
\right|
=
\frac{|h_i|\sqrt{\rho}}{N}
\frac{|\sin(\pi a_i)|}
{\sin\left(\frac{\pi}{N}(1+a_i)\right)}.
\end{equation}
%\subsubsection{Determine the value of $a_i$}
In the following, we use the central sample $k_{\rm c}$ and the stronger neighboring sample, $
\lambda=\max\left\{
\left|y\!\left[k_{\rm l}\right]\right|,
\left|y\!\left[k_{\rm r}\right]\right|
\right\}
$, to find the absolute value of the fractional Doppler $a_i$, where 
$\lambda=\frac{|h_i|\sqrt{\rho}}{N}
\frac{|\sin(\pi a_i)|}
{\sin\left(\frac{\pi}{N}\left(1-|a_i|\right)\right)}.$
%\end{equation}
% Therefore, the ratio between the central sample and the stronger neighbor is
The ratio $\phi=\left|y\!\left[k_{\rm c}\right]\right|/\lambda$ is
\begin{equation}
\phi=
\frac{
\sin\left(\frac{\pi}{N}\left(1-|a_i|\right)\right)
}{
\sin\left(\frac{\pi|a_i|}{N}\right)
}
=
\frac{\sin\left(\frac{\pi}{N}\right)}{\tan\left(\frac{\pi |a_i|}{N}\right)}-\cos\left(\frac{\pi}{N}\right).
\label{eq:phi}
\end{equation}
Therefore, $\phi$ is only a function of $a_i$. 

Solving \eqref{eq:phi} for $\left|a_i\right|$, yields
\begin{equation}
\left|\widehat{a}_i\right|
=
\frac{N}{\pi}
\tan^{-1}
\left[
\frac{
\sin\left(\frac{\pi}{N}\right)
}{
\displaystyle
\phi
+
\cos\left(\frac{\pi}{N}\right)
}
\right].
\end{equation}
 
Finally, we determine the sign of the fractional Doppler by selecting the sign hypothesis that maximizes the correlation between the received signal and the transmitted pilot affected by the estimated delay and candidate Doppler shift.
Let
\begin{equation}
    \widehat{\nu}_i=   \widehat{\alpha}_i+\gamma_i\left|\widehat{a}_i\right|,
    \qquad
    \gamma_i\in\{-1,+1\},
    \label{eq:doppler_sign_hypothesis}
\end{equation}
where $\gamma_i$ denotes the unknown sign of the fractional Doppler shift. 
Let denote the effective channel of the $i^{\rm th}$ path , excluding the path gain, as
\begin{equation}
    \mathbf{H}^i_{\rm eff}
=\mathbf{A}\boldsymbol{\Gamma}_{\mathrm{CPP}_i}\boldsymbol{\Delta}^{{\nu}_i}
    \boldsymbol{\Pi}^{l_i}\mathbf{A}^{\mathrm{H}}.
    \label{eq:candidate_path_operator}
\end{equation}
For each hypothesis $\gamma_i$, inserting $\widehat{l}_i, \widehat{\alpha}_i, \widehat{a}_i$, the corresponding effective channel matrix of $i^{\rm th}$ path is $\widehat{\mathbf{H}}^i_{\rm eff}$.
Consequently, $\gamma_i$ is estimated by maximizing the correlation of the received signal $\mathbf{y}$ and the estimated effective channel matrix $\widehat{\mathbf{H}}^i_{\rm eff}$, i.~e.,
\begin{equation}
\widehat{\gamma}_i
=
\underset{\gamma_i\in\{-1,+1\}}{\arg\max}
\left|
\mathbf{x}_{\rm p}^{\mathrm{H}}
(\widehat{\mathbf{H}}^i_{\rm eff})^{\mathrm{H}}
\mathbf{y}
\right|^{2}.
\label{eq:doppler_sign_estimation}
\end{equation}

\subsection{Path gain estimation}
\label{subsec:path_gain_estimation}

After estimating the delay and Doppler parameters of $i^{\rm th}$ propagation path, the corresponding complex-valued path gain can be obtained by LS estimation. The received signal can be approximated as
\begin{equation}
    \mathbf{y}
    \approx
    h_i\widehat{\mathbf{H}}^i_{\rm eff}\mathbf{x}
    +
    {\boldsymbol{\eta}},
    \label{eq:single_path_gain_model}
\end{equation}
where $h_i$ is the unknown complex path gain and $\boldsymbol{\eta}=[\eta[0],\ldots,\eta[N-1]]^{\rm T}$. The LS estimate of $h_i$ is obtained as
\begin{equation}
    \widehat{h}_i
    =
    \underset{h_i\in\mathbb{C}}{\arg\min}
    \left\|
        \mathbf{y}
        -
        h_i\widehat{\mathbf{H}}^i_{\rm eff}\mathbf{x}_{\rm p}
    \right\|_2^2.
    \label{eq:path_gain_ls_problem}
\end{equation}
Let $\widehat{\mathbf{y}}_{\rm p}^{i}=\widehat{\mathbf{H}}^i_{\rm eff}\mathbf{x}_{\rm p}$, the closed-form solution is given by
\begin{equation}
    \widehat{h}_i
    =
    \frac{
        (\widehat{\mathbf{y}}_{\rm p}^{i})^{\mathrm{H}}\mathbf{y}
    }{
        \|\widehat{\mathbf{y}}_{\rm p}^{i}\|^2
    }
.
    \label{eq:path_gain_ls_solution}
\end{equation}

\subsection{Proposed Multipath Channel Estimation via SIC}

%%%% New description
As explained earlier, our proposed channel estimation technique operates in an iterative manner. Once the parameters of the dominant path are estimated, its contribution to the received signal is reconstructed. 
Since both the pilot and data components associated with the dominant path may interfere with the estimation of the remaining paths, their contributions are removed prior to the next iteration.

As all matrices in \eqref{eq:candidate_path_operator} are unitary, it follows $\left(\mathbf{H}_{\mathrm{eff}}^{i}\right)^{\mathrm -1}\!=\!\left(\mathbf{H}_{\mathrm{eff}}^{i}\right)^{\mathrm H}$.
Therefore, $\left(\mathbf{H}_{\mathrm{eff}}^{i}\right)^{\mathrm H} \mathbf{H}_{\mathrm{eff}}^{i} =\mathbf{I}_N$.
Consequently, the matrix inversion involved in conventional LMMSE detection reduces to a simple scalar normalization, which can be expressed as
\begin{equation}
    \widehat{\mathbf x}_i
    =
    \frac{\widehat{h}_i^{*}}
    {|\widehat{h}_i|^2+\sigma_\eta^2}
    \left(\widehat{\mathbf{H}}_{\mathrm{eff}}^{i}\right)^{\mathrm H}
    \mathbf y_i .
    \label{eq:mmse1}
\end{equation}
For the proposed channel estimation technique, an LMMSE detector is first used to obtain a preliminary estimate of the data symbols, $\widehat{\mathbf{x}}_i$, see line~11 in \textbf{Algorithm~\ref{alg:afdm_pairing_best}}. The contribution of the path $i$ is then reconstructed as $ \widehat{\mathbf{y}}_i=\widehat{h}_i\widehat{\mathbf{H}}^i_{\rm eff}(\mathbf{x}_{\rm p}+\widehat{\mathbf{x}}_i)$. The reconstructed contribution is subsequently removed from the received signal in iteration $i$ to obtain the residual signal
\begin{equation}
\mathbf{y}^{(i+1)}
=
\mathbf{y}^{(i)}-\widehat{\mathbf{y}}_i,
\label{eq:y^(i+1)}
\end{equation}
where $\mathbf{y}^{(0)}=\mathbf{y}$. The resulting signal in \eqref{eq:y^(i+1)} is then used to estimate the remaining dominant path following the steps described in Sections~\ref{subsec:IntDD} to \ref{subsec:path_gain_estimation}. This process is repeated until all the parameters of $P$ paths are estimated. Our proposed SIC-based channel parameter estimation technique is summarized in \textbf{Algorithm~\ref{alg:afdm_pairing_best}}.

\begin{algorithm}[t]
\caption{Proposed Channel Parameter Estimation}
\footnotesize
\label{alg:afdm_pairing_best}
\begin{algorithmic}[1]
\Require $\mathbf{r}$, $\mathbf{x}_{\rm p}$, $\mathbf{s}_{\rm p}$, $P$, $N$, $\mu$ , $\nu_{\max}$ and $l_{\max}$. 
\Ensure Selected parameter set $\mathcal{S}$
    \State Initialize the solution set $\mathcal{S} \gets \emptyset, \mathbf{y}^{(0)}\gets\mathbf{Ar}, \mathbf{r}^{(0)}\gets\mathbf{r}$
\For{$i=0$ to $P-1$}
%\State \textbf{Integer delay and Doppler estimation}
\State $c(n) \gets \left| \left\langle \mathbf{r}^{(i)}, \mathbf{s}_{\rm p}^{(n)} \right\rangle \right|$
\State $n_{1,i}, n_{2,i} \gets$ index of the largest value of $c(n)$ and $|\mathbf{y}^{(i)}|$
\State$ \{\widehat{\alpha}_i, \widehat{l}_i\} \gets \{n_{1,i}, n_{2,i}\}$   
        \State $\widehat{\operatorname{loc}}_i \gets \llrrparen{\alpha_i + 2N c_1 l_i}_N$

        \State $\phi \gets 
        \left| y[\widehat{k}_{\rm c}] \right|/
        \max\left\{\left| y[\widehat{k}_{\rm l}] \right|, \left| y[\widehat{k}_{\rm r}] \right| \right\}
        $
        \State $\widehat{a}_i \gets \dfrac{N}{\pi}
        \tan^{-1}\left(
        \dfrac{\sin\left(\pi/N\right)}
        {\phi+\cos\left(\pi/N\right)}
        \right)$
        \State $\widehat{\gamma}_i \gets \underset{\gamma_i\in\{-1,+1\}}{\arg\max}
        \left|\mathbf{x}_{\rm p}^{\mathrm{H}}
        (\widehat{\mathbf{H}}^i_{\rm eff})^{\mathrm{H}}
        \mathbf{y}^{(i)}
        \right|^{2} \Rightarrow ~\widehat{\nu}_i = \widehat{\alpha}_i + \widehat{\gamma}_i|\widehat{a}_i|$
        \State $\widehat{\mathbf{y}}_{\rm p}^{i}\gets \widehat{\mathbf{H}}^i_{\rm eff}\mathbf{x}_{\rm p}$,     $\widehat{h}_i=\frac{(\widehat{\mathbf{y}}_{\rm p}^{i})^{\mathrm{H}}\mathbf{y}
    }{
        \|\widehat{\mathbf{y}}_{\rm p}^{i}\|^2
    }$
      \State $\widehat{\mathbf x}_i
    \gets
    \frac{\widehat{h}_i^{*}}
    {|\widehat{h}_i|^2+\sigma_w^2}
    \left(\widehat{\mathbf{H}}_{\mathrm{eff}}^{i}\right)^{\mathrm H}
    \mathbf y_i$
        \State $\mathbf{y}^{(i+1)}
            \gets\mathbf{y}^{(i)}-\widehat{h}_i\widehat{\mathbf{H}}^i_{\rm eff}(\mathbf{x}_{\rm p}+\widehat{\mathbf{x}}_i) ~\Rightarrow ~\mathbf{r}^{(i+1)}=\mathbf{A^{\rm H}}\mathbf{y}^{(i+1)}$
        \State $\mathcal{S} \gets \mathcal{S} \cup \{\widehat{h}_i, \widehat{l}_i, \widehat{\nu}_i\}$

\EndFor
\State \Return $\mathcal{S}$
\end{algorithmic}
\end{algorithm}

\section{Computational Complexity Analysis}
In this section, we analyze the computational complexity of our proposed channel parameter estimation technique with the benchmark schemes in terms of the number of complex multiplications (CMs).
The dictionary-based method in~\cite{afdm_journal} consists of two stages: an on-grid search for the integer delay and Doppler parameters, followed by a local search for fractional Doppler refinement.
Accordingly, its computational complexity is in the order of \(\mathcal{O}(N N_l N_{\alpha} N_a)\), where $N_l, N_{\alpha}$ and $N_a$ are dictionary sizes for delay, integer Doppler and fractional Doppler, respectively. The
complexity therefore increases rapidly with the search resolution. Moreover, since the correlation peaks of all paths are detected simultaneously, strong or closely spaced paths may mask weaker ones or each other, respectively.
To avoid this issue, one may think of using the OMP-based method, \cite{omp}, to successively estimate and cancel the propagation paths. At each iteration, correlating the residual signal with all delay-Doppler dictionary atoms requires $\mathcal{O}(N N_l N_{\nu})$ CMs, where $N_{\nu}$ denotes the
size of Doppler dictionary. We use \eqref{eq:mmse1} for the preliminary estimate of the data which has $\mathcal{O}(N )$ CMs. 
Hence, for a $P$-path channel, the resulting complexity is approximately $\mathcal{O}(PNN_lN_\nu)$.

For our proposed technique, the channel parameter estimation is dominated by the time-domain correlation for peak detection and the correlation evaluation for determining the fractional-Doppler sign. Both operations
require $\mathcal{O}(N)$ CMs per path. Including the LMMSE detector of \eqref{eq:mmse1}, the total number of CMs for a $P$-path channel is $\mathcal{O}(3PN)$.

\section{Simulation results}
In this section, we evaluate the performance of the proposed channel estimation scheme. The AFDM system employs \(N=1024\) subcarriers with 4-QAM
modulation. The maximum integer delay and Doppler are set to \(l_{\max}=5\) and \(\nu_{\max}=2\), respectively. The wireless channel consists of \(P=3\) propagation paths and we assume perfect synchronization is achieved.
\begin{figure*}
  \centering
  \begin{minipage}[!t]{0.3\textwidth}
    \includegraphics[scale=0.36]{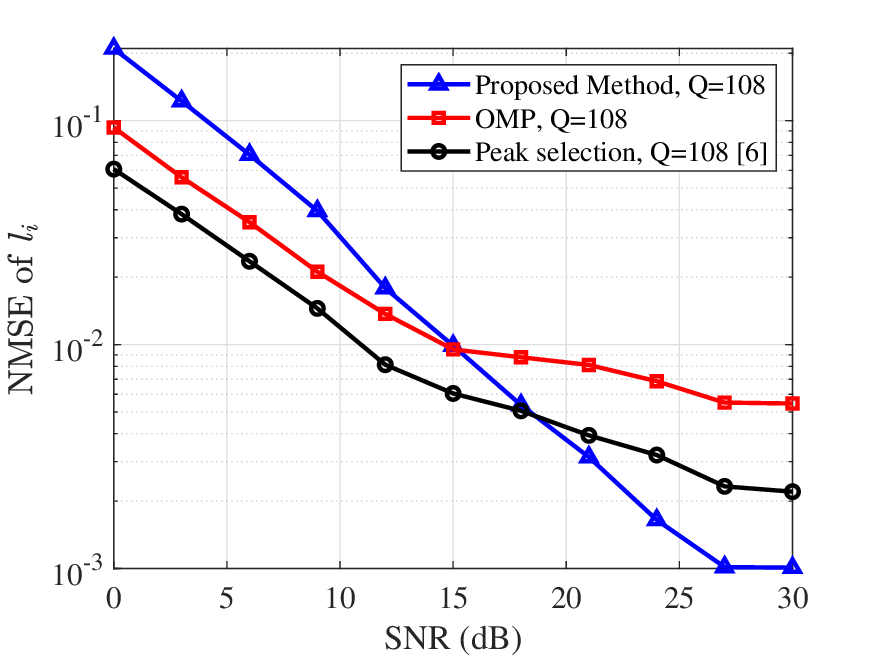}
    \vspace{-0.2cm}
    %\captionsetup{labelfont={color=blue}, textfont={color=blue}}
    \caption{NMSE of delay estimation.}
    \label{fig:delay}
  \end{minipage}
  \vspace{-1em}
  \hspace{1em}
  \begin{minipage}[!t]{0.3\textwidth}
    \includegraphics[scale=0.36]{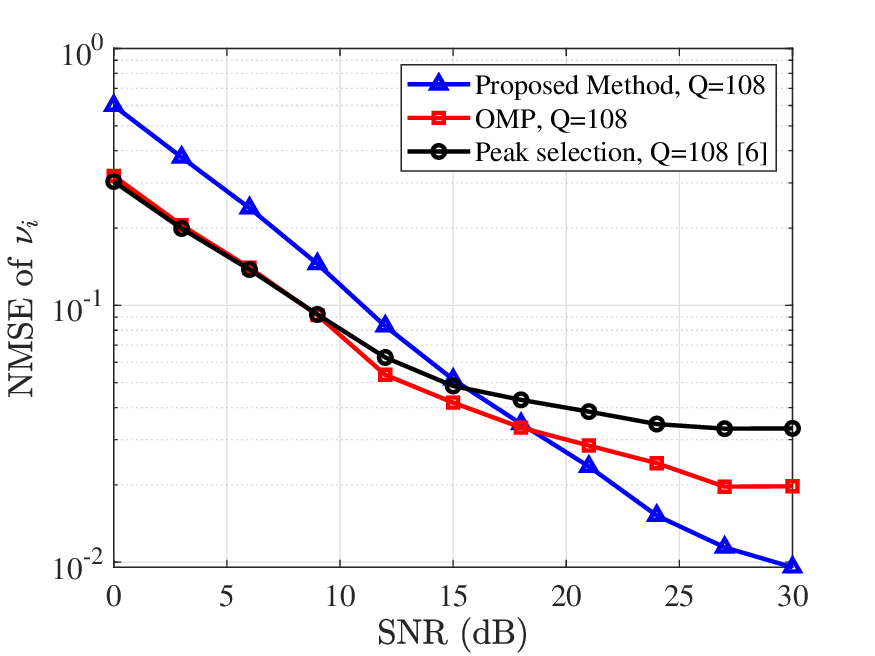}
    \vspace{-0.2cm}
    \caption{NMSE of Doppler estimation.}
    \label{fig:doppler}
    \label{fig:sub22}
  \end{minipage}
  \hspace{1em}
\begin{minipage}[!t]{0.3\textwidth}
    \includegraphics[scale=0.36]{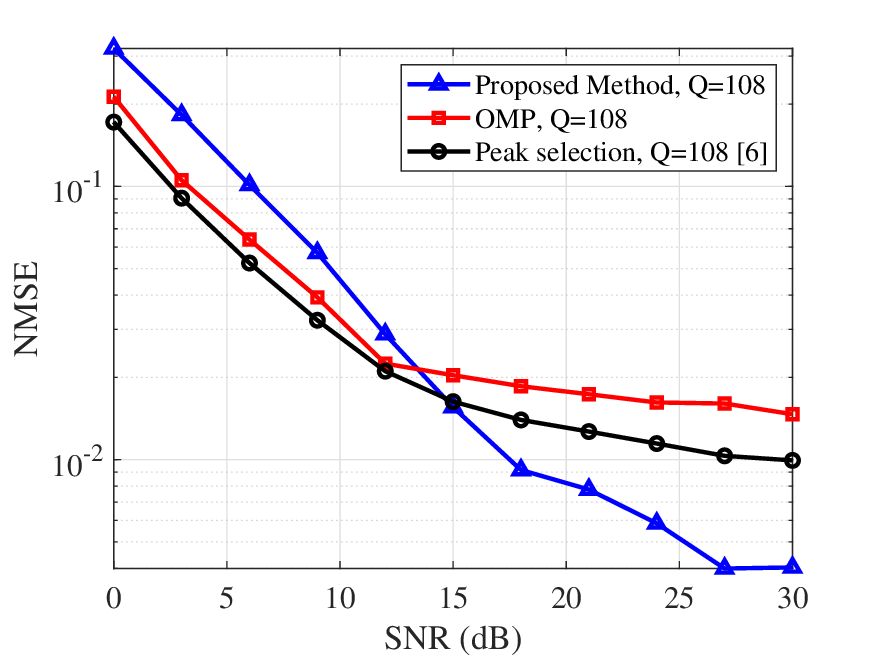}
    \vspace{-0.2cm}
    \caption{NMSE of path gain estimation.}
    \label{fig:gain}
    \label{fig:sub22}
\end{minipage}
  \label{fig:compare_k_n}
\end{figure*}
\begin{figure*}
  \centering
  \begin{minipage}[!t]{0.3\textwidth}
    \includegraphics[scale=0.36]{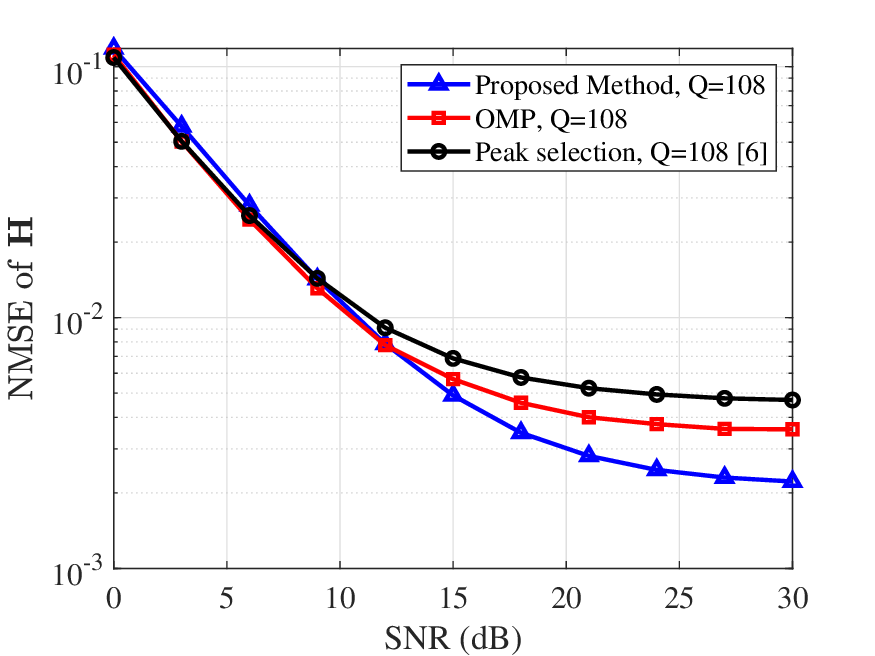}
    \vspace{-0.4cm}
    \caption{NMSE of the channel matrix.}
    \label{fig:nmse}
  \end{minipage}
  %\vspace{-1em}
  \hspace{1em}
  \begin{minipage}[!t]{0.3\textwidth}
    \includegraphics[scale=0.36]{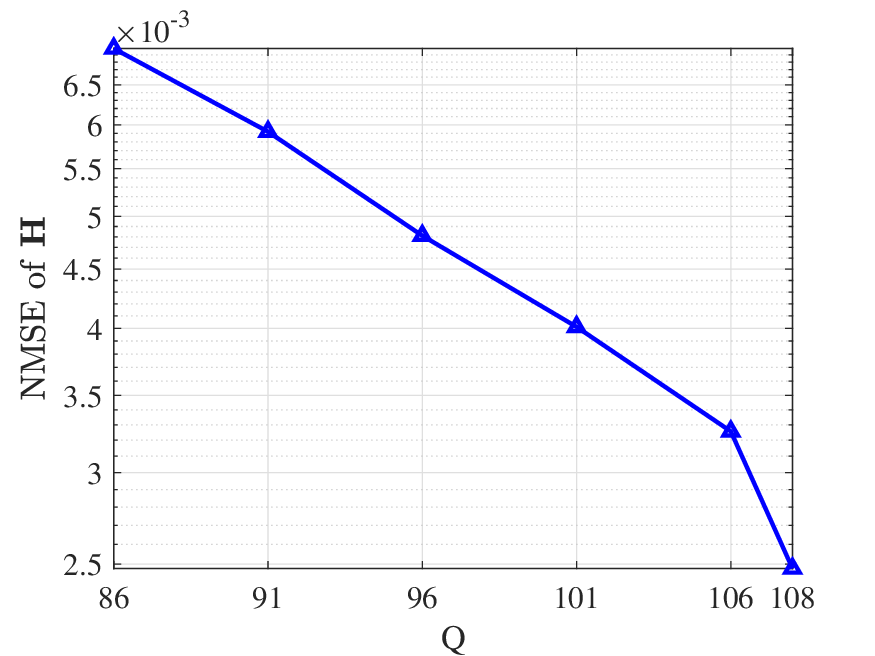}
    \vspace{-0.4cm}
    \caption{Channel matrix NMSE for different $Q$.}
    \label{fig:nmse_Q}
  \end{minipage}
  \hspace{1em}
\begin{minipage}[!t]{0.3\textwidth}
    \includegraphics[scale=0.36]{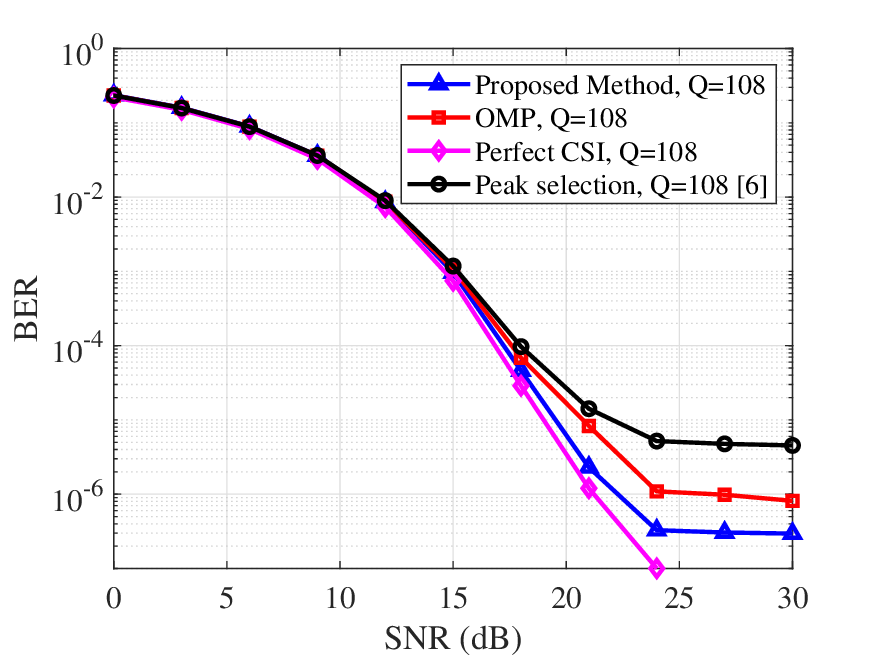}
    \vspace{-0.4cm}
    \caption{BER performance.}
    \label{fig:ber}
\end{minipage}
  \vspace{-0.5cm}
\end{figure*}
A line-of-sight (LOS)-dominant Rician tapped-delay-line (TDL) channel model is considered\cite{3gpp}. The first tap corresponds to the LOS component and
is modeled as a Rician fading channel, whereas the remaining taps are modeled as independent Rayleigh
fading. The average tap powers follow an exponentially decaying power-delay profile (PDP) and are normalized according to
$\sum_{\ell=0}^{P-1} P_{\ell}=1,$
where \(P_{\ell}\) denotes the average power of the \(\ell\)-th
propagation path.
The LOS channel coefficient is given by
\begin{equation}
    h_0
    =
    \sqrt{P_0}
    \left(
        \sqrt{\frac{K_0}{K_0+1}}e^{j\phi_0}
        +
        \sqrt{\frac{1}{K_0+1}}g_0
    \right),
\end{equation}
where \(K_0\) is the Rician \(K\)-factor, \(\phi_0\) is uniformly
distributed over \([0,2\pi)\), and
\(g_0\sim\mathcal{CN}(0,1)\). 
In the simulations, the delay and Doppler grid resolutions for the peak selection method and the OMP-based benchmark are set to \(1\) and \(0.01\), respectively. Under these settings, the proposed method requires approximately \(300\) times and \(800\) times
fewer CMs than the peak selection and OMP-based methods, respectively, demonstrating a substantial reduction in computational complexity.

We evaluate the channel-parameter estimation performance in Fig.~\ref{fig:delay}, \ref{fig:doppler}, and \ref{fig:gain}. 
From these figures, we can have following observations. 
Firstly, our scheme achieves lower parameter normalized mean square error (NMSE) when the signal-to-noise ratio (SNR) is higher than \(15\)~dB compared to the benchmarks. This improvement is mainly attributed to the direct estimation of the fractional Doppler component in the proposed method, which alleviates the limited resolution introduced by the discretized Doppler dictionary.

Fig.~\ref{fig:nmse} presents the NMSE of the reconstructed
channel matrix. It can be observed that the proposed method achieves lower NMSE when SNR is above \(15\)~dB compared to OMP and peak selection methods.
In the low-SNR region, weak paths may lead to unreliable peak detection for \eqref{eq:n1} and \eqref{eq:fa_mapping},
%, causing \eqref{eq:affine_shift} and \eqref{eq:fa_mapping} to yield no valid solution and
resulting in missed path estimates. Under this scenario, the estimation error of delay, Doppler, and path gain are set to $l_{\max}/2$, $\nu_{\max}$, and square root of the average path power, respectively. This penalization contributes to the higher parameter estimation NMSE at low SNRs. However, since the missed paths are weak, their contributions to the overall channel are limited. The resulting channel-matrix NMSE remains comparable to that of the benchmark methods at low SNRs.

To evaluate the potential of the proposed method for reducing guard
overhead, we examine the channel estimation NMSE with a reduced guard
size $Q$ at an SNR of \(25\)~dB. As shown in Fig.~\ref{fig:nmse_Q},
when $Q$ is reduced to $96$, the proposed method achieves an NMSE of
approximately $4.8\times10^{-3}$, which is around the NMSE achieved by the peak selection method at the same SNR. This corresponds to a \(11\%\) reduction in guard overhead while providing a comparable channel estimation accuracy.

Fig.~\ref{fig:ber} compares the BER performance of the
different channel estimation schemes using the LMMSE detector. We observe that the proposed method outperforms OMP and peak selection methods and achieves a lower error floor. These results indicate that the proposed method offers not only lower computational complexity, but also less overhead, better channel estimation accuracy, and detection reliability. For reference, the BER obtained using perfect channel state information (CSI) is also included. The proposed methods are nearly identical to the perfect-CSI benchmark when the SNR is lower than \(18\)~dB, while only a small performance gap is observed at higher SNRs.
\vspace{-0.2cm}
\section{Conclusion}
%\vspace{-0.2cm}
This paper proposed a low-complexity impulse-pilot-based channel parameter estimation method for AFDM systems by exploiting the inherent delay-Doppler shift properties and AFDM input-output relationship.
The integer delay and Doppler parameters are estimated from the shift properties in the time and affine domains, while the fractional Doppler is obtained from the Doppler-induced leakage around the impulse pilot.
For multipath channels, we introduced a novel iterative SIC framework to successively estimate, reconstruct, and cancel the dominant paths.
Simulation results show that the proposed method achieves lower
computational complexity, improved channel estimation and BER
performance, and reduced pilot overhead compared with the benchmark
solutions.

%%%%%%%%%%%%%%%%%%%%%%%%%
\bibliographystyle{IEEEtran} 
\bibliography{reference}

\end{document}